# Alteration of skeletal muscle energy metabolism assessed by 31P MRS in clinical routine, part 2: Clinical application

Antoine Naëgel[1,2], Hélène Ratiney[1], Jabrane Karkouri[1,2,3], Djahid Kennouche[1,4], Nicolas Royer[1,4], Jill M. Slade[5], Jérôme Morel[6], Pierre Croisille[1,7], Magalie Viallon[1,7]

[1] Université de Lyon, INSA-Lyon, Université Claude Bernard Lyon 1, UJM-Saint Etienne, CNRS, Inserm, CREATIS UMR 5220, U1206, Lyon, France, [2] Siemens Healthcare SAS, Saint-Denis, France, [3] Wolfson Brain Imaging Center, University of Cambridge, Cambridge, United Kingdom, [4] LIBM - Laboratoire Interuniversitaire de Biologie de la Motricité, [5]Department of Radiology, Michigan State University, East Lansing, USA ; [6]anaesthetics and intensive care department, UJM-Saint-Etienne, Centre Hospitalier Universitaire de Saint-Étienne, Saint-Etienne, France ; [7] Radiology department, UJM-Saint-Etienne, Centre Hospitalier Universitaire de Saint-Étienne, Saint-Etienne, France

Abbreviations: ADP, adenosine diphosphate; ATP, adenosine triphosphate; BMI, body mass index; CK, creatine kinase; CV, coefficient of variation; EDSS, expanded disability status scale; ICU, intense care unit; $k_{PCr}$, rate constant of post-exercise PCr recovery; MS, multiple sclerosis; MVC, maximum voluntary contraction; $H_2PO_4^-$, diprotonated phosphate; PCr, phosphocreatine; Pi, inorganic phosphates; QCS, quality control score; QUEST, QUantitation based on QUantum ESTimation; $R_{ind}$, individual correction factor; $R_{fixe}$, fixed correction factor; $\tau_{PCr}Rec$, time constant of post-exercise PCr recovery; $\tau_{Pi}Rec$, time constant of post-exercise Pi recovery; $\tau_{PCr}Ex$, time constant of exercise PCr depletion; $\tau_{Pi}Ex$, time constant of exercise Pi depletion; TCr, total creatine; $Vi_{PCr}$, The initial rate of PCr recovery; $V_{max}$, the maximum aerobic capacity.

Corresponding author: Magalie Viallon, Université de Lyon, INSA-Lyon, Université Claude Bernard Lyon 1, UJM-Saint Etienne, CNRS, Inserm, CREATIS UMR 5220, U1206, Lyon, France.

Email : magalie.viallon@creatis.insa-lyon.fr

**Abstract Summary**

Background: In this second part of a two-part paper, we intend to demonstrate the impact of the previously proposed advanced quality control pipeline. To understand its benefit and challenge the proposed methodology in a real scenario, we chose to compare the outcome when applying it to the analysis of two patient populations with a significant but highly different types of fatigue: COVID19 and multiple sclerosis (MS).

Experimental: $^{31}$P-MRS was performed on a 3T clinical MRI, in 19 COVID19 patients, 38 MS patients, and 40 matched healthy controls. Dynamic acquisitions using an MR-compatible ergometer ran over a rest(40s), exercise(2min), and a recovery phase(6min). Long and short TR acquisitions were also made at rest for $T_1$ correction. The advanced data quality control pipeline presented in part 1 is applied to the selected patient cohorts to investigate its impact on clinical outcomes. We first used power and sample size analysis to estimate objectively the impact of adding QCS. Then, comparisons between patients and healthy control groups using validated QCS were performed using unpaired T-tests or Mann-Whitney tests ($p<0.05$).

Results: The application of the QCS resulted in increased statistical power, changed the values of several outcome measures, and reduced variability (SD). A significant difference was found between the $T_{1PCr}$ and $T_{1Pi}$ of MS patients and healthy controls. Furthermore, the use of a fixed correction factor led to systematically higher estimated concentrations of PCr and Pi than when using individually corrected factors. We observed significant differences between the two patient populations and healthy controls for resting [PCr] – MS only, [Pi], [ADP], [$H_2PO_4^-$] and pH – COVID19 only, and post-exercise [PCr],[Pi] and [$H_2PO_4^-$] - MS only. The dynamic indicators $\tau_{PCr}$, $\tau_{Pi}$, $Vi_{PCr}$ and $V_{max}$ were reduced for COVID19 and MS patients compared to controls.

Conclusion: Our results show that QCS in dynamic $^{31}$P-MRS studies results in smaller data variability and therefore impacts study sample size and power. Although QCS resulted in discarded data and therefore reduced the acceptable data and subject numbers, this rigorous and unbiased approach allowed for proper assessment of muscle metabolites and metabolism in patient populations. The outcomes include an increased metabolite $T_1$, which directly affect the T1 correction factor applied to the amplitudes of the metabolite, and a prolonged $\tau_{PCr}$ indicating reduced muscle oxidative capacity for patients with MS and COVID19.

**Introduction**

As previously mentioned in part 1, evaluating skeletal muscle energetics and metabolism is of clinical interest to monitor neuromuscular or cardiovascular degenerative diseases and quantify muscle fatigability[1–3].

Implementing a standardized $^{31}$P-MRS dynamic acquisition protocol to evaluate skeletal muscle energy metabolism and monitor muscle fatigability [1,2], while being compatible with various longitudinal clinical studies on diversified patient cohorts, requires a high level of technicality and expertise. These difficulties led us to develop a specific quality control pipeline to objectively identify and highlight outliers and standardize the data analysis. In this second part of a two-part paper, we intend to demonstrate the impact of the previously proposed advanced quality control pipeline. To test its added value and challenge the proposed methodology in a real scenario, we chose to compare the outcome when applying it to the analysis of 2 patient populations with significant but different etiology of fatigue: COVID19 and multiple sclerosis (MS).

Indeed, fatigue is a common but subjective and multifactorial symptom in many patients including those with MS and patients with advanced COVID19 infection[4]. Following COVID19 infection, various respiratory and neuromuscular disorders have been described that manifest in a decrease in physical capacities[5]. Muscle damage in COVID19 is the result of a multitude of interrelated factors, caused by infection, deconditioning, drug cytotoxicity or malnutrition, as discussed in detail in recent reviews[6,7]. Many sources of dysfunction can lead to fatigue, ranging from the central nervous system through local control of muscle activation (i.e. central fatigue) to mitochondrial respiration (i.e. peripheral fatigue) [8]. In MS patients, previous studies investigating energy metabolism using $^{31}$P-MRS have shown that fatigue is likely related to impaired excitation-contraction coupling and abnormal energy metabolism [9–11]. Moreover, it has been shown that muscle fatigability and mitochondrial activity are closely related, with cytoplasmic Pi accumulation and ATP production dynamics being the main mechanisms potentially involved in muscle fatigue [12,13] in short and intensive exercises.

The objective of this second paper is to evaluate the impact of QCS on clinical cohorts with a precise clinical question to demonstrate the clinical impact and the pertinence of the proposed approach. The aims of this second part are first to explore the impact of the quality control routine from part 1 on clinical results. The second aim is to explore the impact of using patient specific T1 estimates versus fixed T1 values from the literature. The final aim was to provide comprehensive results concerning the energy metabolism of populations subject to chronic fatigue, in particular in patients with advanced COVID19.

**Experimental**

*Studied populations*

This work was conducted on 19 COVID19 patients (age: 64±13 years, Body Mass Index (BMI): 25.98±4.28 kg/m$^2$), 38 MS patients (age: 45±9 years, BMI: 25.26±4.58 kg/m$^2$), and 40 matched healthy controls (described below). Participants were recruited and provided written informed consent. All experimental procedures, inclusion and exclusion criteria were approved by the Institutional Review Boards (CPP Nord Ouest VI, ethics committee agreement 19.02.22.52507 received on November 21, 2019), approved by the ethics committee (Protection Personnes Ile de France VIII, #20 04 05), registered at ClinicalTrials.gov (#NCT04363606) and agreed with the principles in the Declaration of Helsinki. Healthy controls for COVID19 and MS were selected based on age and BMI of the patient groups; 10 controls matched the COVID19 patients (age: 59±6 years, BMI: 24.55±3.64 kg/m$^2$) and 35 controls matched the MS patients (age: 45±11 years, BMI: 24.81±3.78 kg/m$^2$). The healthy control subjects are a subsample of controls from part 1.

The COVID19 patients presented severe forms of the disease, resulting in hospitalization and an ICU admission with mechanical ventilation. COVID19 patients spent 33.1 ± 18.6 days in intensive care, including 25.3 ± 16.4 days on mechanical ventilation. ICU patients are described here with the Sepsis-related Organ Failure Assessment (SOFA) score[14] and the Simplified Acute Physiology Score (SAPS II)[15], which respectively determine and monitor the status of a patient with organ failure and measure the severity of illness in patients admitted to ICUs. COVID19 patients had a SOFA score of 5.9±2.4 and a SAPS II score of 35.1±9.7. These scores are associated with predicted mortality rates of 20%[15,16]. The MS patients included in the study were diagnosed and had clinical signs for more than 2 years (13±4 years) and a disability rating scale (Expanded Disability Status Scale, EDSS[17]) score lower than 5. A score of 0 represents a normal neurological condition, a score of 5 represents a disability profound enough to interfere with the activity of a normal day and a score of 10 is MS-related death.

Changes in the demographics are provided for comparisons before and after a comprehensive quality control scoring (QCS; see Supplementary Material Table 1). Care was taken to quantify and statistically test for group differences after QCS since the matching between healthy controls and patients was made prior to QCS. This could be a limitation of the work if the group demographics changed after initial group matching.

*MR experiments and exercise protocol*

Spectroscopy and NMR imaging were performed on a 3T MR unit (MAGNETOM PRISMA, Siemens Healthineers, Germany). The sequence, hardware, and exercise protocol as well as the data analysis and $^{31}$P-MRS Quantification have been described in part 1 (ref). Briefly, non-localized MRS-FID sequence[18] was used to acquire: 1) resting $^{31}$P-MRS acquisitions before the exercise, with TR=30sec (12 acquisitions) and TR=4sec (32 acquisitions) to obtain a fully-relaxed and a $T_1$ saturated spectrum of the quiescent muscles[19] respectively; 2) a series of $^{31}$P-MRS acquisitions running over a 40s rest phase (10 acquisitions), a 2min exercise phase (30 acquisitions) and a final 6min recovery phase (90 acquisitions). $^{31}$P dynamic MRS acquisitions were synchronized with the periodic plantar flexions on the ergometer every 4s, presented to the subject by visual stimuli (e-prime (version 2.0.10.261), Psychological Software Tools (Sharpsburg, PA, USA).

*Saturation correction factor and $T_1$ estimation per metabolites*

$T_1$ was estimated within MATLAB R2021a (The MathWorks, Inc., Natick, Massachusetts, United States), for each metabolite and each subject, from fully-relaxed and $T_1$ saturated spectra, using monoexponential fitting[19]. The correction factor $R_i$ derived from the $T_1$ measurement is:

$$R_i = \frac{1}{1 - e^{-TR/T_{1i}}} \quad (Eq1)$$

With i referring to the considered metabolite: phosphocreatine (PCr), inorganic phosphate (Pi), α-, β-, and γ-Adenosine Tri Phosphate (ATP).

Individually (personalized) estimated $T_{1i}$ was further used to correct the metabolite amplitudes measured in the subsequent dynamic acquisitions, performed at short TR, hence in a saturation regime.

The impact of individualized T1 was measured by comparing it with a correction factor derived from a fixed T1 value in the literature[20].

*Data analysis and $^{31}P$-MRS quantification*

$^{31}$P-MRS Quantification was performed as described in part 1, combining steps in MATLAB and using the QUEST (QUantitation based on QUantum ESTimation) method in its command-line version[21]. PCr, Pi, α-, β-, and γ-ATP amplitudes, were estimated throughout the protocol and extracted to distinguish three phases: rest, exercise and recovery. The concentration of the metabolites PCr, Pi, ATP, ADP, and the pH were extracted at rest, post-exercise and after recovery, as well as the concentration of diprotonated phosphate, $[H_2PO_4^-]$[22,20]. The PCr depletion (percentage) during plantar exercise, the time constants of PCr

($\tau_{PCr}$Ex and $\tau_{PCr}$Rec) and Pi ($\tau_{Pi}$Ex and $\tau_{Pi}$Rec) were also derived during exercise (ex) and recovery (rec). The initial recovery rate of PCr ($Vi_{PCr}$) and the maximum aerobic capacity ($V_{max}$) were derived according to [23,24]. The $V_{max}$ is based on a model selected to consider the enzymatic aspect, through the constant Km and the concentration of [ADP], considered a key regulator in the oxidative synthesis of ATP[25]. Furthermore, the $Vi_{PCr}$ allows access to the contractile cost generated by a muscular effort. Finally, the resting PCr/Pi ratio was calculated, which is known to be a strong indicator of the state of oxidative function and the predominant muscle fiber type in the muscle studied[20]. Concerning the dynamic indicators, $Vi_{PCr}$ and $V_{max}$ were estimated only for the exercise phase. As described in part 1, the QC pipeline will evaluate QC variables (PCr depletion, R2 $\tau_{PCr}$Rec |$\tau_{Pi}$Rec, Outliers, $\tau_{PCr}$Ex| $\tau_{Pi}$Ex , R2 $\tau_{PCr}$Ex|$\tau_{Pi}$Ex, $\tau_{PCr}$Rec|$\tau_{Pi}$Rec), each associated with a QC score (0 to -3). The resulting scores objectively guide decisions on when to accept or reject data.

All the quantitative variables were calculated with and without the advanced quality control described in Part 1 to demonstrate the outcome of the proposed methodology in a realistic scenario with a clear clinical question, This is the most direct approach to evaluate the impact of QCS on the clinical outcomes and obtain indicators on the power of the technique.

*Statistical analysis*

Continuous variables were first screened for normality using the Shapiro-Wilk test and were reported as the mean ± standard deviation (SD) except otherwise stated. Comparisons between patients and control groups were performed using unpaired T-tests or Mann-Whitney tests.

We used power and sample size analysis to estimate the impact of adding QCS when performing comparisons between control and patient groups using $\tau_{PCr}$Rec, in COVID-19 and MS populations. Power calculations were performed using two-sample means Satterthwaite's t-test assuming unequal variances, α set to 5%, and using the descriptive statistics (μ,σ) measured in patients and matched controls as input.

While the number of control subjects was set as fixed, the sample size of patients was iteratively tested in the range (2, 100).

We also used the measures of τPCrRec obtained in the current study, to perform an a priori power analysis and compute the sample size required with or without QCS. Power (1-β) was set to 80% and α to 5% (two-sided Satterthwaite's t test assuming unequal variances), expecting an equal-group allocation. Statistical analyses were performed using Xlstat (version 2022.3.1) and Stata 17 (Statacorp, College Station,Tx, USA). A p value <0.05 was considered to indicate a statistically significant difference.

**Results**

*Impact of the Adaptive Analysis and Quality Control*

Table 1 summarizes patient $^{31}$P-MRS muscle quantitative metabolic variables as well as differences against their matched controls and QC variables at rest, during exercise and during recovery for both cohorts. Table 2 provides the estimates when the same markers are obtained without using QCS, while Table 3 provides the estimates obtained with QCS but using fixed $T_1$ values. The comparison of the derived estimates and the difference in the statistical analysis illustrates the impact of the refinements of the proposed analysis method on the results.

The results of the quality control based on the rating scale are summarized in Figure 1. For the COVID19 cohorts, QCS excluded the exercise part of the protocol for 21% of patients and 10% of controls. For the MS cohort, QCS excluded exercise data from 11% of patients and 6% of controls. Correction on the first point of the recovery and the corresponding adjustment was required in 3 COVID19 patients and 1 MS patient whose exercise part could not be exploited (i.e. the subject's QCS score was between -2 and -3). Among these subjects, where the first point of the recovery was detected as corrupted, an increase in R2 was observed for 2 of them. The proportion of subjects whose data was totally or partially excluded (only exercise) was lower in healthy controls than in patients. The QCS resulted in the exclusion of all behavior data from 21% of COVID19 patients and 10% of their control subjects. Similarly, for MS cohorts, the QCS excluded all behavior data from 21% of MS patients and 8% of their control subjects.. The proportion of subjects with acceptable data across the dynamic protocol (i.e., score of zero) is 32% for COVID19 patients and 20% for controls, 39% for MS patients and 29% for controls. Overall, the use of the QCS resulted in the automatic classification of 53% of COVID19 and 60% of MS patients.

Additionally, the impact of the application of our quality control was investigated by exploiting the results with and without QCS (Table 1 vs Table 2). Without QCS, all patients and controls were incorporated in the analysis. The repetition of the analysis with and without quality control results in differences in the values, data variability (SD) and significance level (p-value) of the extracted variables for the comparison of our populations. The quality control indicators are logically negatively impacted, for example, the R2 coefficient corresponding to the adjustment to obtain the $\tau_{Pi}Ex$ for COVID19 patients with QCS is 0.89±0.08 and without QCS is 0.68±0.31. The SD was reduced across several measures in both patient groups and healthy controls across several metabolites as well as $\tau_{PCr}$ and $\tau_{Pi}$. Smaller SD were observed despite smaller sample sizes.

Figure 2 shows the results of the power analysis in MS and COVID19 patients using $\tau_{PCr}Rec$ parameter results. In both COVID19 and MS patient groups, we observed that there was a clear benefit in using QCS-controlled data compared to results obtained without QCS, since QCS always translated to increased power (1-β), those lowering type II error (false negative). An a priori power analysis was performed to compute the sample size requirements for an expected power = 80% and α =5%. We found that in the case of MS patients with QCS applied, the total sample size requirement would be 44 subjects (22 per group) to detect a difference of 7.62 (effect size). In comparison without using QCS, the total sample size would increase to 52 subjects (26 per group), for a smaller effect size (2.33). In the case of COVID19 patients with QCS applied, the total sample size requirement would be 34 subjects (17 per group) to detect a difference of 10.88, while without QCS, the total sample size would increase to 74 subjects (37 per group) to detect a difference of 8.32. Overall, power analysis based on τPCrRec COVID data patients showed that sample size could be reduced by 54% when using QCS (compared to without). Hereafter, the analysis of the groups by age and BMI are presented after QCS. After using the QCS on patients and controls, BMI and age remained similar to the original groups (see Supplementary Material Table 1).

*Impact of using patient specific $T_1$ estimates versus fixed $T_1$ values*

Our work also analyzed the influence of using a personalized estimated $T_1$ correction factor per metabolite instead of a fixed factor. The use of a fixed correction factor ($R_{fixe}$) led to systematically higher estimated concentrations of PCr and Pi than when using individually corrected factors ($R_{ind}$) (see Figure 3 and Table 3). In MS controls, [PCr]$_{rest}$ values were significantly different when calculated with a personalized or a fixed correction factor: [PCr]$_{rest}$ with individual $T_1$ correction was 33.51±4.26 mM while [PCr]rest with default literature $T_1$ correction was 36.04±4.46 mM.

The variation in the amplitude of PCr and Pi due to the application of the correction factor has an impact on the calculation of $Vi_{PCr}$, $V_{max}$ and the PCr/Pi ratio rest and this has an impact on the significant differences. The correlation and Bland-Altman charts of [PCr]$_{rest}$ and [Pi]$_{rest}$ with individual and fixed *T1* correction (see Figure 4) indicate that the values obtained for the parameters estimated with a fixed T1 are higher than with an individualized *T1*. Indeed, for these parameters, the p-value increases with the application of an individual correction factor: PCr/Pi rest ratio (with $R_{ind}$ p=0.020 and with $R_{fixe}$ p=0.015); $Vi_{PCr}$ and $V_{max}$ (with $R_{ind}$ p=0.014 and with $R_{fixe}$ p=0.009) for MS cohorts (Table 1 and Table 3). The application of an average value of the personalized T1 as a correction factor for each population was also investigated; Supplementary Material Table 2 summarizes the observed values for [PCr] and [Pi] at rest.

*Detailed clinical outcomes*

Figures 5 and 6 display the distribution of a selection of the main physiological markers of interest obtained with our validated QCS procedure. The primary outcome of $\tau_{PCr}$Rec revealed longer time constants for the patient groups relative to their controls (p≤0.029). For patients versus their matched control groups, there was a 33% slower $\tau_{PCr}Rec$ for COVID19 (p=0.029, see Table 1 and Figure 6) and a 23% slower $\tau_{PCr}Rec$ for MS (p=0.001, see Table 1 and Figure 5). On the other hand, there was a *decrease* in $Vi_{PCr}$ of 35% for COVID19 (p=0.003) and 18% for MS (p=0.014); and in $V_{max}$ of 35% for COVID19 (p=0.004) and 16% for MS (p=0.014). In addition, resting muscle metabolic measures were decreased in the patient groups including PCr/Pi (COVID19 p≤0.023 and MS p≤0.020).

Focusing on the MS cohorts, patients had increased PCr and Pi concentration for each phase of the protocol (rest, exercise and recovery) compared to controls (p≤0.001). In addition, there is a significant *increase* of the $T_1$ of PCr and Pi by 7% and 9% (p≤0.035), respectively. Focusing on the COVID19 cohorts, patients had increased Pi concentration and pH for the rest phase compared to controls (p≤0.035). For more details regarding the results please refer to Table 1, which compiles all the results for the two populations studied.

Despite a significant difference at rest regarding pH between groups (COVID19 cohorts), there was no significant change in pH during exercise nor severe acidosis. The change in pH during exercise in patients and controls in COVID19 were 0.041 ± 0.032 and 0.021±0.031, respectively and in MS were 0.022±0.017 and 0.021±0.018, respectively. The muscular exercise led to a significant depletion of PCr, with ~41% hydrolysis for COVID19 cohorts and ~35% for MS cohorts. Therefore the depletion of PCr reached the level of expectation to create a significantly detectable change (>20%) while preserving the pH[20], warrantying a fair comparison of $\tau_{PCr}$ and $\tau_{Pi}$ between the studied populations after QCS.

The MVC measured for COVID19, MS patients and controls were respectively 564.3±175.7, 507.5±162.9 and 537.1±144.8 N.m$^{-1}$, with no significant differences between patient groups and their respective control groups (p=0.545 for MS, and p=0.632 for COVID19). During the $^{31}$P-MRS exercise, the %MVC was 90.3±10.2% for COVID19 patients, 87.2±13% for MS patients and 91.7±11.7% for the combined control groups, with no significant differences between patient groups and their respective control groups (p=0.58 for MS, and p=0.884 for COVID19). These checks attest that the data have been acquired under favorable conditions and can support the analysis and the results.

## Discussion

*Detailed discussion of the Adaptive Analysis and Quality Control*

The results obtained from the comparison of populations with and without the use of quality control (Table 1 vs. Table 2) reveal the impact of data corruption during the exercise and its consequences on the results. These results support the importance of systematically incorporating data quality control as determined by its impact within our protocols. The introduced quality control method allowed us to detect the corrupted data efficiently and to preserve as much data as possible by considering the exercise and recovery parts separately resulting in the most optimal and conservative data exclusion. In light of the above-mentioned QC results, the implementation of a data quality control methodology is essential to objectively assess data quality and make valid comparisons and interpretations.

Quality control should exclude as few patients as possible from the study without bias, preserving as much of the data as possible. This implies creating quality indicators that are as accurate as possible, targeting only corrupted data and considering the individual features of each pathology investigated. In this regard, the indicators should be derived from the experience of the operators and validated by the expert community. Only one of our QCS procedures is the not fully automatic and deserves enhanced discussion: the reselection of the first point for fitting the metabolite recoveries. The same data quality control was applied to all subjects with four subjects requiring reselection (3 COVID19, 1 MS). In our case, the proposed reselection of the starting point of the recovery fit is unlikely to introduce a bias given the monoexponential pattern expected with mild exercise (see part 1 for further comment). The selection of a highly oxidative muscle group (plantar flexors), along with the mild exercise intensity of the current study (0.25 Hz) and small pH changes support the use of a monoexponential pattern notwithstanding an oversimplification of the initial rate of PCr resynthesis following exercise including a glycolytic component[26,27].

*Detailed discussion of the impact of using patient specific $T_1$ estimates versus fixed $T_1$ values*

This study highlighted a significant difference between measured $T_1$ values and $T_1$ values reported in the literature for PCr and Pi metabolites of the MS group. More specifically, the $T_{1PCr}$ and $T_{1Pi}$, in the control groups in both cohorts shows not only a significant difference with their respective patients' group but also a reduced $T_1$ relaxation time compared to the literature.

The measurement of $T_1$ of metabolites requires an additional measurement, with a long TR acquisition, which considerably increases the protocol length. In most studies, it is generally established that the $T_1$ of metabolites does not change and is used in the same way for all corrections[10,22,28–30]. The increase in $T_1$ of metabolites (PCr and Pi) with pathology observed here in both study populations challenges such an assumption and supports the choice of a protocol including a long TR measurement to take into account the $T_1$ weighting of the metabolites in an individualized way. In a further approach, the application of an average value of the personalized T1 as a correction factor for each population was used, although results are not systematically different compared to the use of a fixed T1 from the literature. Note that this observation cannot necessarily be generalized to other populations and supports the time taken during the protocol to perform this measurement prior to using any assumption.

*Detailed discussion of the clinical outcomes*

The primary findings indicate a significant prolongation of the recovery period observed in both patient populations, along with reduced $Vi_{PCr}$ and $V_{max}$. This shows reduced mitochondrial function in both patient populations. The mild exercise selected in the study yielded no muscle pH differences during exercise and no significant muscle acidification while attaining substantive PCr hydrolysis yielding a robust design. Further analysis failed to detect significant differences between the two patient populations as reported in the appendix.

The processes of ATP production during exercise remain complex to explain due to the many metabolic pathways used simultaneously by the body. The recovery phase in this design is reflective of the oxidative pathway. In the two studied patient populations, we observed a significant increase of $\tau_{PCr}Rec$ or $\tau_{Pi}Rec$, as well as a substantial decrease of $Vi_{PCr}$ and $V_{max}$. These coupled indicators point to an alteration in mitochondrial respiration. The return of PCr to equilibrium is governed by creatine kinase activity at the expense of oxidative resynthesis of ATP. Our results confirm that the model also permits the determination of $V_{max}$ from the resting concentration of PCr. These indicators are considered insensitive to exercise intensity and metabolic conditions at the end of exercise (acidosis), hence reliable and complementary indicators to $\tau_{PCr}$. These indicators are useful when evaluating and monitoring diseases that may lead to higher acidosis in patients. In the case of serious acidosis, other mechanisms come into play and can disturb the return to equilibrium of the PCr and complicate the analysis of the $\tau_{PCr}Rec$, which would no longer be the reflection of the oxidative mechanism.

These results are consistent with prior findings showing that MS is associated with reduced oxidative capacity [9] and in the present study using a mild exercise that may be better tolerated by patients with fatigue. These are among the first data on in vivo skeletal muscle mitochondrial function following COVID19 infection showing the infection is associated with reduced muscle oxidative capacity. In both patient groups, the primary outcomes may be associated with the significant deconditioning effect due to their diseases as well as consequences (immobilization, sedentary lifestyle, etc.). This may result in an impact on muscle and metabolism that affect $^{31}$P-MRS measurements [5,9–11,31]. The effect of ageing, which can be associated with deconditioning, on muscular capacity has revealed reductions in oxidative capacity with age [31,32]. However, it is important to note that when physical activity levels are accounted for and matched in comparative studies, oxidative capacity is maintained well into older age [33].

The COVID19 patient population was comprised of older and older middle-aged adults who likely suffered from both physical and psychological consequences of the disease for which the severity required treatment in the intensive care unit. While no formal viral typing is available, most of these patients experienced early variants of the virus, which tended to be more virulent and at a time when vaccines and treatments were still in development. These factors may have contributed to reduced muscle function and contributed to the ability to execute the exercise resulting in more corrupted data due to motion during data acquisition. Analysis of patients with new variants will be necessary to fully understand COVID19 effects on muscle metabolism as well as follow-up studies to examine changes following COVID19 recovery including impacts of long COVID on muscle metabolism. Despite several papers highlighted in a review on COVID19 impacts on skeletal muscle, to the best of the authors knowledge, limited data on COVID19 muscle oxidative capacity or related fatigue data were available for comparison.

For MS patients, no significant difference in the dynamic indicators, $\tau_{PCr}Ex$ and $\tau_{Pi}Ex$, was observed during the exercise phase. However, higher concentrations of PCr and Pi were observed in the patients at rest and at the end of exercise (post), reflecting an accumulation of these metabolites in the cellular environment. ADP concentration in MS patients is significantly higher at rest but this difference was not observed post-exercise. This shows an increase in the dynamics of ADP utilization in MS patients. A higher mobilization of ADP reveals an intensification of ATP production and thus an increase in workload, which is usually accompanied by a change in pH. However, we did not observe a significant change in pH during exercise. The outcomes from other studies have been mixed on exercise pH changes in MS [10,11], with one study showing no acidification following exercise compared to a control group [9].

For COVID19 patients, no significant difference in the dynamic indicators, $\tau_{PCr}Ex$ and $\tau_{Pi}Ex$, was observed during the exercise phase. However, there was a significant increase in Pi concentration at rest, which was not accompanied by an increase in Pi concentration post-exercise. Similarly, the concentration of ADP seems to follow this pattern. These variations are in favour of a disturbing dynamic concerning Pi and ADP. Like MS patients, COVID19 patients seem to mobilize ATP production to a greater extent during the exercise cycle.

**Conclusion**

In this second part, the study shows the impact of the advanced quality control pipeline on clinical populations. The application of QCS increases statistical power and provides more robust and precise results. This quality control approach helps the operator to focus on problematic cases and to get the most out of datasets by classifying data of nearly 50% of the subjects included in the cohorts. Overall, the study demonstrates an alteration in muscle metabolism in MS and COVID19 patients compared to controls. The results reveal a significant influence of the correction of metabolite amplitude by an individual relaxation scale factor, which supports the argume51nt for incorporating an appropriate resting TR measure into clinical protocols. In conclusion, in the context of the translation of a methodological protocol to clinical studies, points of vigilance are mandatory throughout the process, from acquisition to quantification.

**Acknowledgments**

The authors thank the subjects for their participation in this study. We thank Guillaume Y. Millet for his expertise in exercise physiology and involvement in this project. This work was partly supported by the LABEX PRIMES (ANR-11-LABX-0063), Siemens Healthineers and Jabrane Karkouri was supported by the European Union's Horizon 2020 research and innovation programme under grant agreement No 801075.

**Data Availability Statement**

The data that support the findings of this study are available on request from the corresponding author. The data are not publicly available due to privacy or ethical restrictions.

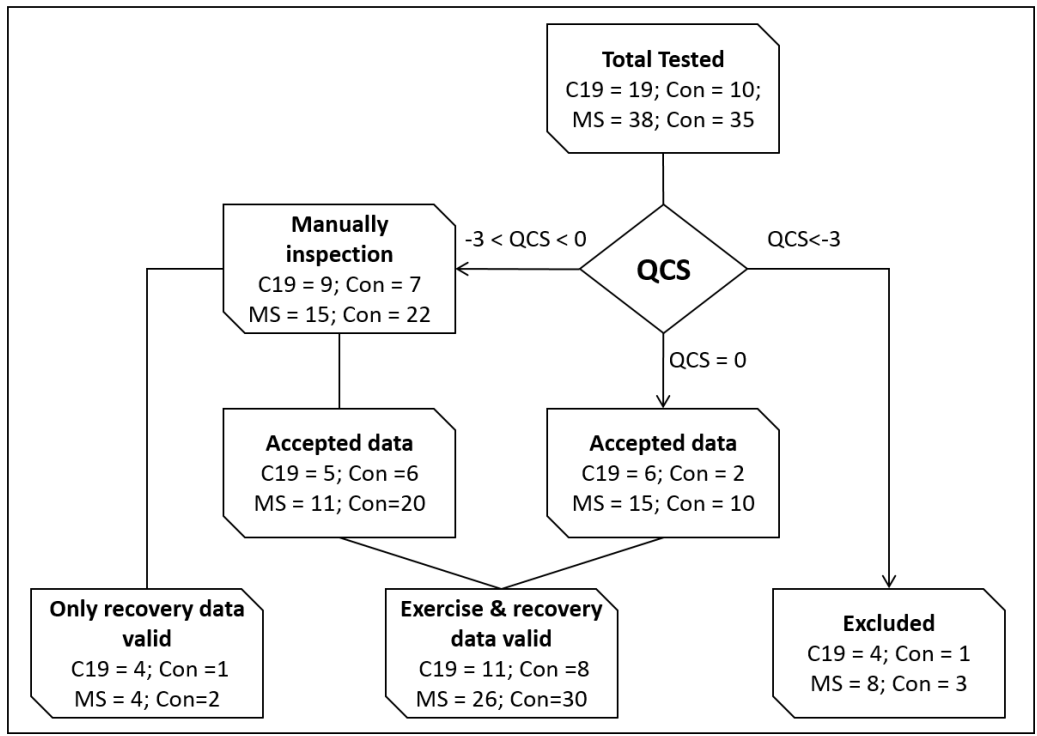

Figure 1 : Flowchart of the application of the quality control score. Details of the individual analysis of subjects and results in support of decision-making. C19, number of COVID19 patient's; MS, number of MS patient's; Con, number of matched controls with COVID19 and MS patients.

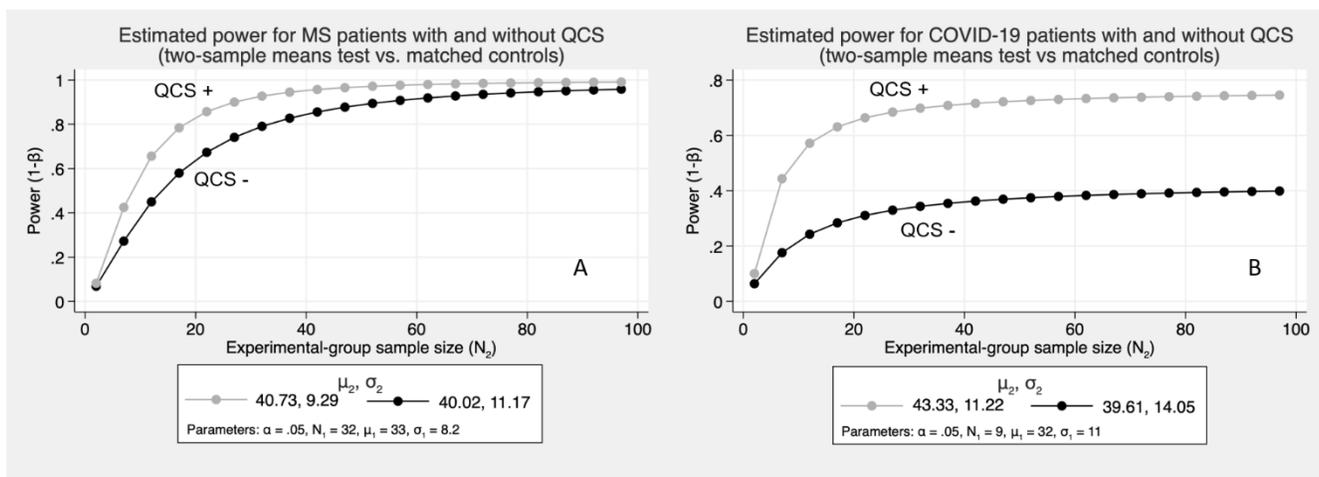

Figure 2: Power analysis curves for $\tau_{PCr}$Recovery in (A) MS and (B) COVID19 patients. Calculations are using two-sample means Satterthwaite's t test assuming unequal variances, using the results collected in controls and corresponding patients with or without QCS (see Tables 1 and 2) while sample size of experimental groups varies.

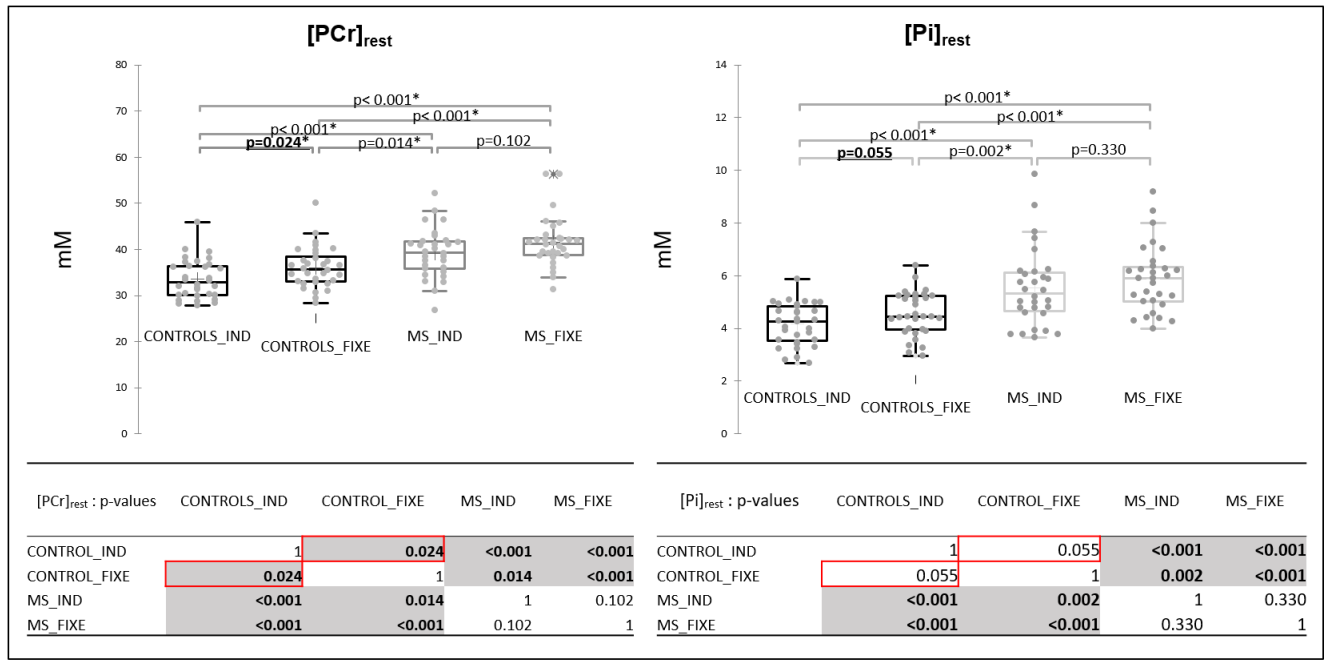

Figure 3: Boxplots and table describing the resting [PCr] and [Pi] obtained on the MS cohorts with individual correction factor and with fixed correction factor. A significant difference is observed in MS controls between $[PCr]_{rest}$ with individual correction factor and $[PCr]_{rest}$ with fixed correction factor. *T-test with $p<0.05$.

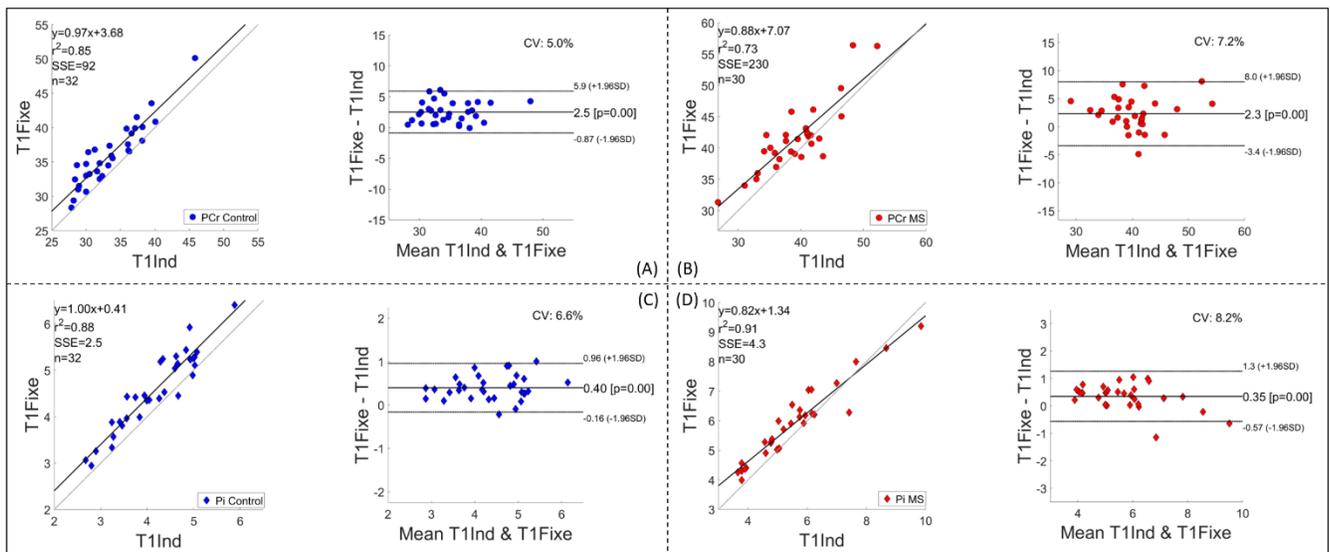

Figure 4: Correlation and Bland-Altman plots of $[PCr]_{rest}$ (A,B) and $[Pi]_{rest}$ (C,D) estimated with a T1 individualized and fixed obtained on MS patients (B,D) and their matched controls (A,C). Comparably good correlation for the two parameters was found with R2 = 0.85 for PCr Control, R2 = 0.73 for PCr MS, R2 = 0.88 for Pi Control and R2 = 0.91 for Pi MS. The values obtained by a fixed T1 are higher than those obtained by an individualized T1 ($p < 0.05$).

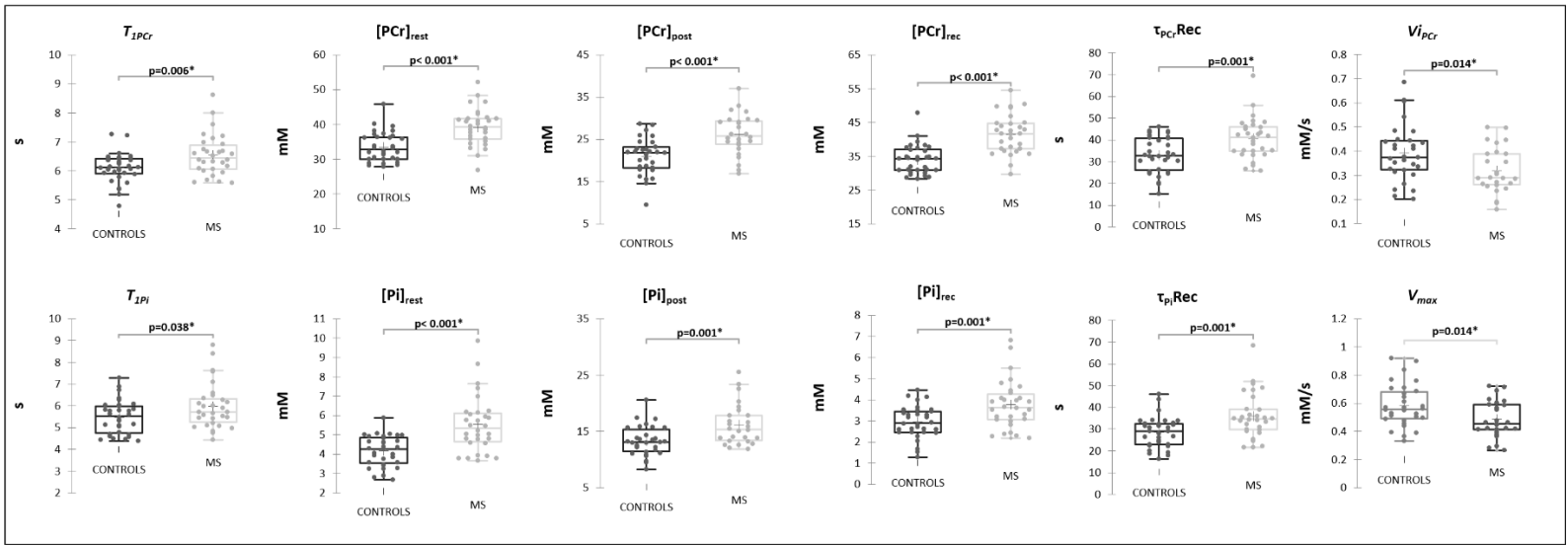

Figure 5: Boxplots of the feature of interest obtained on the MS cohorts. *T-test with Statistical significance set to p<0.05.

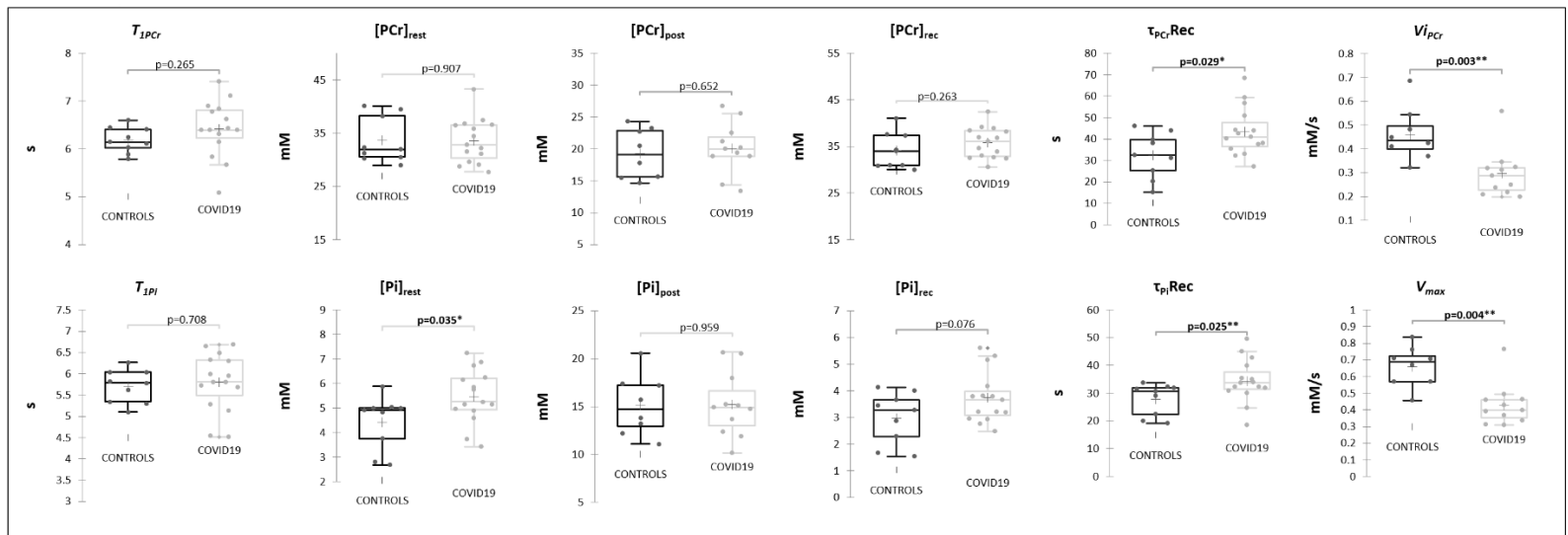

Figure 6: Boxplots of the feature of interest obtained on the COVID19 cohorts. *T-test with Statistical significance set to p<0.05. **Mann-Whitney test with Statistical significance set to p<0.05.

| Rest | COVID19 | | | | MS | | | |
|---|---|---|---|---|---|---|---|---|
| | Patients (N=15) | Controls (N=9) | P-values** | Variation | Patients (N=30) | Controls (N=32) | P-values** | Variation |
| $T_{1PCr}$ (s) | 6.42 (0.59) | 6.18 (0.27) | 0.265 | | 6.55 (0.71) | 6.10 (0.51) | **0.006** | ↗ |
| $T_{1Pi}$ (s) | 5.80 (0.68) | 5.70 (0.39) | 0.708 | | 5.97 (1.05) | 5.48 (0.77) | **0.038** | ↗ |
| $T_{1\gamma ATP}$ (s) | 5.27 (0.58) | 5.32 (0.43) | 0.821 | | 5.41 (0.57) | 5.14 (0.65) | 0.092 | |
| $T_{1\alpha ATP}$ (s) | 3.39 (0.22) | 3.39 (0.17) | 0.987 | | 3.43 (0.44) | 3.26 (0.24) | 0.062 | |
| $T_{1\beta ATP}$ (s) | 3.46 (0.61) | 3.37 (0.34) | 0.693 | | 4.23 (2.73) | 3.34 (0.38) | 0.074 | |
| $[PCr]_{rest}$ (mM) | 33.59 (4.21) | 33.68 (4.32)* | 0.907 | | 39.20 (5.32) | 33.51 (4.26) | **<0.001** | ↗ |
| $[Pi]_{rest}$ (mM) | 5.47 (1.10) | 4.43 (1.10) | **0.035** | ↗ | 5.54 (1.47) | 4.16 (0.79) | **<0.001** | ↗ |
| $[ATP]_{rest}$ (mM) | 7.96 (0.51) | 8.14 (0.55)* | 0.482 | | 8.08 (0.58) | 8.21 (0.43) | 0.351 | |
| $[ADP]_{rest}$ (µM) | 15.21 (3.43) | 11.07 (4.98) | **0.024** | ↗ | 14.14 (2.81) | 11.16 (3.37) | **<0.001** | ↗ |
| $[H_2PO_4^-]_{rest}$ (mM) | 1.70 (0.37) | 1.44 (0.34) | 0.092 | | 1.79 (0.48) | 1.35 (0.24) | **<0.001** | ↗ |
| $pH_{rest}$ | 7.10 (0.02) | 7.07 (0.01) | **0.003** | ↗ | 7.07 (0.03) | 7.07 (0.03) | 0.473 | |
| PCr/Pi rest | 6.36 (1.40) | 7.97 (1.82) | **0.023** | ↘ | 7.38 (1.48) | 8.26 (1.41) | **0.020** | ↘ |

| Exercise | COVID19 | | | | MS | | | |
|---|---|---|---|---|---|---|---|---|
| | Patients (N=11) | Controls (N=8) | P-values** | Variation | Patients (N=26) | Controls (N=30) | P-values** | Variation |
| $\tau_{PCr}Ex$ (s) | 26.66 (9.41)* | 34.21 (20.85)* | 0.492 | | 34.03 (14.13) | 36.38 (18.91) | 0.605 | |
| $\tau_{Pi}Ex$ (s) | 27.59 (10.12) | 25.59 (8.92) | 0.661 | | 31.56 (19.09) | 25.86 (10.31) | 0.162 | |
| $[PCr]_{post}$ (mM) | 20.10 (4.02) | 19.26 (3.89) | 0.652 | | 26.17 (4.81) | 20.96 (4.31) | **<0.001** | ↗ |
| $[Pi]_{post}$ (mM) | 15.22 (3.37) | 15.14 (3.16) | 0.959 | | 16.19 (3.65) | 13.33 (2.65) | **0.001** | ↗ |
| $[ATP]_{post}$ (mM) | 7.27 (1.25)* | 7.84 (0.82)* | 0.206 | | 7.77 (0.60) | 7.94 (0.62) | 0.312 | |
| $[ADP]_{post}$ (µM) | 54.85 (9.10) | 64.86 (28.22) | 0.282 | | 47.86 (8.79) | 55.87 (24.09) | 0.115 | |
| $[H_2PO_4^-]_{post}$ (mM) | 5.02 (1.35) | 4.93 (1.11) | 0.879 | | 5.19 (1.15) | 4.29 (0.90) | **0.002** | ↗ |
| $pH_{post}$ | 7.06 (0.04) | 7.07 (0.04)* | 0.395 | | 7.07 (0.03) | 7.07 (0.03) | 0.892 | |
| pH min | 6.93 (0.11)* | 6.88 (0.16) | 0.717 | | 6.97 (0.05) | 6.93 (0.12) | 0.190 | |
| pH change | 0.04 (0.03) | 0.02 (0.03)* | 0.351 | | 0.02 (0.02) | 0.02 (0.02) | 0.868 | |
| PCr depletion (%) | 39.51 (9.69) | 42.27 (14.84) | 0.629 | | 32.49 (7.45) | 37.41 (12.54) | 0.086 | |
| PCr change (mM) | 13.05 (3.12) | 14.73 (6.45) | 0.461 | | 12.55 (3.15) | 12.74 (4.89) | 0.865 | |
| $Vi_{PCr}$ (mM/s) | 0.30 (0.10)* | 0.46 (0.11) | **0.003** | ↘ | 0.32 (0.09) | 0.39 (0.12) | **0.014** | ↘ |
| $V_{max}$ (mM/s) | 0.43 (0.13)* | 0.66 (0.12) | **0.004** | ↘ | 0.49 (0.13) | 0.58 (0.15) | **0.014** | ↘ |
| QC outcomes | | | | | | | | |
| R2 $\tau_{PCr}Ex$ | 0.93 (0.03) | 0.93 (0.06)* | 0.395 | | 0.93 (0.03) | 0.93 (0.06) | 0.745 | |
| R2 $\tau_{Pi}Ex$ | 0.89 (0.08)* | 0.92 (0.06) | 0.545 | | 0.91 (0.05) | 0.91 (0.06) | 0.790 | |

| Recovery | COVID19 | | | | MS | | | |
|---|---|---|---|---|---|---|---|---|
| | Patients (N=15) | Controls (N=9) | P-values** | Variation | Patients (N=30) | Controls (N=32) | P-values** | Variation |
| $\tau_{PCr}Rec$ (s) | 43.33 (11.22) | 32.45 (10.66) | **0.029** | ↗ | 40.73 (9.29) | 33.11 (8.24) | **0.001** | ↗ |
| $\tau_{Pi}Rec$ (s) | 34.39 (7.74) | 27.83 (5.71)* | **0.025** | ↗ | 36.08 (10.49) | 28.55 (7.09) | **0.001** | ↗ |
| $[PCr]_{rec}$ (mM) | 35.82 (3.30) | 34.12 (3.82) | 0.263 | | 41.56 (5.78) | 34.03 (4.32) | **<0.001** | ↗ |
| $[Pi]_{rec}$ (mM) | 3.73 (0.96) | 2.98 (0.97) | 0.076 | | 3.77 (1.14) | 2.94 (0.76) | **0.001** | ↗ |
| $[ATP]_{rec}$ (mM) | 8.20 (0.12) | 8.20 (0.16) | 0.976 | | 8.22 (0.13) | 8.21 (0.14) | 0.789 | |
| $[ADP]_{rec}$ (µM) | 9.76 (0.86) | 9.44 (0.78) | 0.379 | | 9.58 (0.54) | 9.57 (1.07) | 0.946 | |
| $[H_2PO_4^-]_{rec}$ (mM) | 1.25 (0.37)* | 1.02 (0.33) | 0.347 | | 1.28 (0.39) | 1.00 (0.25) | **0.001** | ↗ |
| $pH_{rec}$ | 7.05 (0.03) | 7.03 (0.03) | 0.120 | | 7.04 (0.02) | 7.04 (0.03) | 0.667 | |
| PCr repletion (%) | 48.57 (15.20)* | 43.51 (13.22) | 0.519 | | 36.64 (7.94) | 38.82 (11.09) | 0.381 | |
| PCr change (mM) | 17.32 (5.18) | 15.10 (5.67) | 0.336 | | 15.25 (4.13) | 13.27 (4.34) | 0.070 | |
| QC outcomes | | | | | | | | |
| R2 $\tau_{PCr}Rec$ | 0.93 (0.03) | 0.93 (0.08)* | 0.558 | | 0.93 (0.05) | 0.94 (0.05) | 0.681 | |
| R2 $\tau_{Pi}Rec$ | 0.88 (0.07) | 0.91 (0.05) | 0.269 | | 0.91 (0.06) | 0.91 (0.06) | 0.873 | |
| CV of $\tau_{PCr}Rec$ (%) | 5.40 (3.53) | 3.92 (1.97) | 0.262 | | 4.13 (2.18) | 5.02 (3.62) | 0.251 | |
| CV of $\tau_{Pi}Rec$ (%) | 5.97 (5.45)* | 6.33 (3.28) | 0.263 | | 4.03 (3.10) | 6.08 (2.67) | **0.007** | ↘ |

Table 1: COVID19, multiple sclerosis (MS) cohorts, and matched summary results with the Quality Control Score (QCS) and personalized T1 values . Results are descriptive statistics : mean (SD). * The Shapiro-Wilk test has a value of p<0.05, the comparison between the populations is therefore made by a Mann-Whitney test with a statistical significance set at p<0.05. **T-test with Statistical significance set to p<0.05. The variation indicates the direction of trends for the patient group compared to the control group. "Change" refers to the difference between the resting and post-exercise state (exercise phase) and to the difference between the post-exercise and post-recovery state (recovery phase).

| Rest | COVID19 | | | | MS | | | |
|---|---|---|---|---|---|---|---|---|
| | Patients (N=19) | Controls (N=10) | P-values** | Variation | Patients (N=38) | Controls (N=35) | P-values** | Variation |
| $T_{1PCr}$ (s) | 6.48 (0.56) | 6.24 (0.32) | 0.229 | | 6.49 (0.69) | 6.14 (0.51) | **0.017** | ↗ |
| $T_{1Pi}$ (s) | 5.84 (0.71) | 5.65 (0.41) | 0.434 | | 5.93 (0.96) | 5.49 (0.74) | **0.033** | ↗ |
| $T_{1\gamma ATP}$ (s) | 5.33 (0.58) | 5.30 (0.41) | 0.904 | | 5.40 (0.54) | 5.15 (0.63) | 0.077 | |
| $T_{1\alpha ATP}$ (s) | 3.36 (0.23) | 3.36 (0.18) | 0.953 | | 3.41 (0.41) | 3.26 (0.24) | 0.058 | |
| $T_{1\beta ATP}$ (s) | 3.58 (0.70) | 3.36 (0.32) | 0.368 | | 4.05 (2.45) | 3.34 (0.36) | 0.093 | |
| $[PCr]_{rest}$ (mM) | 33.68 (4.66) | 34.20 (4.39)* | 0.604 | | 38.21 (5.34) | 33.51 (4.22) | **<0.001** | ↗ |
| $[Pi]_{rest}$ (mM) | 5.46 (1.11) | 4.47 (1.05)* | **0.024** | ↗ | 5.35 (1.46) | 4.17 (0.77) | **<0.001** | ↗ |
| $[ATP]_{rest}$ (mM) | 7.74 (1.29)* | 8.17 (0.53)* | 0.456 | | 8.02 (0.60) | 8.19 (0.43) | 0.161 | |
| $[ADP]_{rest}$ (µM) | 14.31 (4.39) | 10.92 (4.72) | 0.065 | | 13.94 (2.80) | 11.26 (3.30) | **<0.001** | ↗ |
| $[H_2PO_4^-]_{rest}$ (mM) | 1.71 (0.36) | 1.45 (0.33) | 0.065 | | 1.73 (0.47) | 1.35 (0.23) | **<0.001** | ↗ |
| $pH_{rest}$ | 7.09 (0.03)* | 7.07 (0.01) | **0.024** | ↗ | 7.07 (0.03) | 7.07 (0.02) | 0.676 | |
| PCr/Pi rest | 6.37 (1.33) | 7.97 (1.72) | **0.010** | ↘ | 7.48 (1.56) | 8.22 (1.36) | **0.035** | ↘ |

| Exercise | COVID19 | | | | MS | | | |
|---|---|---|---|---|---|---|---|---|
| | Patients (N=19) | Controls (N=10) | P-values** | Variation | Patients (N=38) | Controls (N=35) | P-values** | Variation |
| $\tau_{PCr}$Ex (s) | 28.14 (11.83) | 97.63 (209.04)* | 0.604 | | 458.99 (2606.82) | 56.97 (112.78) | 0.365 | |
| $\tau_{Pi}$Ex (s) | 32.54 (39.32)* | 32.68 (29.00)* | 0.735 | | 1053.88 (6258.80) | 927.18 (5313.39) | 0.926 | |
| $[PCr]_{post}$ (mM) | 18.81 (8.70) | 20.37 (5.52) | 0.612 | | 26.36 (6.15) | 21.02 (4.77) | **<0.001** | ↗ |
| $[Pi]_{post}$ (mM) | 12.81 (4.80) | 14.75 (3.28) | 0.264 | | 14.28 (5.25) | 13.01 (2.90) | 0.209 | |
| $[ATP]_{post}$ (mM) | 6.22 (2.21)* | 7.92 (0.75)* | **0.016** | ↘ | 7.55 (1.35) | 7.85 (1.10) | 0.309 | |
| $[ADP]_{post}$ (µM) | 228.50 (857.44)* | 61.20 (27.61) | 0.910 | | 45.66 (17.40) | 54.69 (23.26) | 0.063 | |
| $[H_2PO_4^-]_{post}$ (mM) | 4.01 (1.87) | 4.74 (1.18) | 0.270 | | 4.55 (1.66) | 4.16 (1.03) | 0.237 | |
| $pH_{post}$ | 7.12 (0.27)* | 7.08 (0.04) | 0.839 | | 7.08 (0.03) | 7.08 (0.05) | 0.950 | |
| pH min | 6.92 (0.18)* | 6.87 (0.18) | 0.456 | | 6.98 (0.06) | 6.93 (0.13) | 0.074 | |
| pH change | 0.11 (0.23) | 0.03 (0.03) | 0.278 | | 0.02 (0.02) | 0.03 (0.03) | 0.654 | |
| PCr depletion (%) | 45.68 (21.83) | 39.96 (15.54) | 0.469 | | 30.99 (14.45) | 37.04 (13.15) | 0.066 | |
| PCr change (mM) | 14.86 (6.15)* | 13.83 (6.27)* | 0.573 | | 11.85 (5.51) | 12.49 (4.91) | 0.605 | |
| $Vi_{PCr}$ (mM/s) | 0.64 (1.16)* | 0.44 (0.11) | 0.266 | | 0.36 (0.41) | 0.40 (0.16) | 0.608 | |
| $V_{max}$ (mM/s) | 0.75 (1.05)* | 0.64 (0.12) | 0.164 | | 0.54 (0.49) | 0.59 (0.20) | 0.529 | |
| QC outcomes | | | | | | | | |
| R2 $\tau_{PCr}$Ex | 0.87 (0.14)* | 0.91 (0.10)* | 0.115 | | 0.89 (0.10) | 0.91 (0.09) | 0.386 | |
| R2 $\tau_{Pi}$Ex | 0.68 (0.31)* | 0.89 (0.11)* | **0.035** | ↗ | 0.80 (0.35) | 0.89 (0.13) | 0.160 | |

| Recovery | COVID19 | | | | MS | | | |
|---|---|---|---|---|---|---|---|---|
| | Patients (N=19) | Controls (N=10) | P-values** | Variation | Patients (N=38) | Controls (N=35) | P-values** | Variation |
| $\tau_{PCr}$Rec (s) | 39.61 (14.05) | 31.29 (10.70) | 0.114 | | 40.02 (11.17) | 32.02 (8.68) | **0.001** | ↗ |
| $\tau_{Pi}$Rec (s) | 38.12 (11.49)* | 28.10 (5.45)* | **0.001** | ↗ | 38.47 (12.63) | 28.88 (7.27) | **<0.001** | ↗ |
| $[PCr]_{rec}$ (mM) | 36.31 (4.83) | 34.53 (3.83) | 0.321 | | 40.5 (5.70) | 34.06 (4.21) | **<0.001** | ↗ |
| $[Pi]_{rec}$ (mM) | 4.14 (1.33) | 3.05 (0.94) | **0.030** | ↗ | 3.81 (1.07) | 2.99 (0.74) | **<0.001** | ↗ |
| $[ATP]_{rec}$ (mM) | 8.17 (0.20)* | 8.19 (0.16) | 0.875 | | 8.23 (0.13) | 8.20 (0.14) | 0.326 | |
| $[ADP]_{rec}$ (µM) | 10.18 (1.32)* | 9.53 (0.79) | 0.115 | | 9.63 (0.58) | 9.63 (1.05) | 0.988 | |
| $[H_2PO_4^-]_{rec}$ (mM) | 1.34 (0.41) | 1.04 (0.32) | **0.050** | ↗ | 1.29 (0.36) | 1.01 (0.24) | **<0.001** | ↗ |
| $pH_{rec}$ | 7.06 (0.04) | 7.04 (0.03) | 0.066 | | 7.04 (0.03) | 7.04 (0.03) | 0.786 | |
| PCr repletion (%) | 48.52 (21.20) | 40.64 (15.43) | 0.309 | | 34.61 (14.22) | 38.19 (12.36) | 0.256 | |
| PCr change (mM) | 17.5 (8.40)* | 14.15 (6.13) | 0.308 | | 14.14 (6.19) | 13.04 (4.64) | 0.395 | |
| QC outcomes | | | | | | | | |
| R2 $\tau_{PCr}$Rec | 0.86 (0.18)* | 0.90 (0.12)* | 0.403 | | 0.89 (0.12) | 0.92 (0.08) | 0.208 | |
| R2 $\tau_{Pi}$Rec | 0.80 (0.19)* | 0.90 (0.06) | 0.104 | | 0.86 (0.12) | 0.89 (0.07) | 0.217 | |
| CV of $\tau_{PCr}$Rec (%) | 8.79 (13.74)* | 5.07 (4.09)* | 0.668 | | 7.49 (12.87) | 5.60 (4.13) | 0.409 | |
| CV of $\tau_{Pi}$Rec (%) | 6.70 (5.98)* | 6.08 (3.19) | 0.512 | | 5.44 (6.43) | 6.39 (3.10) | 0.432 | |

Table 2: COVID19, multiple sclerosis (MS) cohorts, and matched summary results without using the Quality Control Score (QCS). Results are descriptive statistics : mean (SD) . * The Shapiro-Wilk test has a value of p<0.05, the comparison between the populations is therefore made by a Mann-Whitney test with a statistical significance set at p<0.05. **T-test with Statistical significance was set to p<0.05. The variation indicates the direction of trends for the patient group compared to the control group. The values with bold font indicate a substantial difference in SD (larger sample variation) compared to data with QCS reported in Table 1. Smaller SD were observed despite smaller sample sizes which tend to decrease variation.

| Rest | COVID19 | | | | MS | | | |
|---|---|---|---|---|---|---|---|---|
| | Patients (N=15) | Controls (N=9) | P-values** | Variation | Patients (N=30) | Controls (N=32) | P-values** | Variation |
| $T_{1PCr}$ (s) *** | 6.60 | 6.60 | | | 6.60 | 6.60 | | |
| $T_{1Pi}$ (s) *** | 6.10 | 6.10 | | | 6.10 | 6.10 | | |
| $T_{1\gamma ATP}$ (s) *** | 5.00 | 5.00 | | | 5.00 | 5.00 | | |
| $T_{1\alpha ATP}$ (s) *** | 3.00 | 3.00 | | | 3.00 | 3.00 | | |
| $T_{1\beta ATP}$ (s) *** | 3.70 | 3.70 | | | 3.70 | 3.70 | | |
| [PCr]$_{rest}$ (mM) | 35.39 (3.81) | 36.69 (4.06)* | 0.519 | | 41.52 (5.46) | 36.04 (4.46) | **<0.001** | ↗ |
| [Pi]$_{rest}$ (mM) | 5.84 (1.01) | 4.83 (1.18) | **0.035** | ↗ | 5.88 (1.27) | 4.56 (0.84) | **<0.001** | ↗ |
| [ATP]$_{rest}$ (mM) | 7.96 (0.51) | 8.14 (0.55)* | 0.482 | | 8.08 (0.58) | 8.21 (0.43) | 0.351 | |
| [ADP]$_{rest}$ (µM) | 15.21 (3.43) | 11.07 (4.98) | **0.024** | ↗ | 14.14 (2.81) | 11.16 (3.37) | **<0.001** | ↗ |
| [H$_2$PO$_4^-$]$_{rest}$ (mM) | 1.82 (0.34) | 1.57 (0.37) | 0.099 | | 1.90 (0.42) | 1.48 (0.25) | **<0.001** | ↗ |
| pH$_{rest}$ | 7.10 (0.02) | 7.07 (0.01) | **0.003** | ↗ | 7.07 (0.03) | 7.07 (0.03) | 0.473 | |
| PCr/Pi rest | 6.21 (1.14) | 8.03 (2.13) | **0.012** | ↘ | 7.25 (1.20) | 8.11 (1.47) | **0.015** | ↘ |

| Exercise | COVID19 | | | | MS | | | |
|---|---|---|---|---|---|---|---|---|
| | Patients (N=11) | Controls (N=8) | P-values** | Variation | Patients (N=26) | Controls (N=30) | P-values** | Variation |
| $\tau_{PCr}Ex$ (s) | 26.66 (9.41)* | 34.21 (20.85)* | 0.492 | | 34.03 (14.13) | 36.38 (18.91) | 0.605 | |
| $\tau_{Pi}Ex$ (s) | 27.59 (10.12) | 25.59 (8.92) | 0.661 | | 31.56 (19.09) | 25.86 (10.31) | 0.162 | |
| [PCr]$_{post}$ (mM) | 21.35 (4.45) | 20.82 (4.18) | 0.793 | | 27.73 (4.82) | 22.47 (4.65) | **<0.001** | ↗ |
| [Pi]$_{post}$ (mM) | 16.08 (2.54) | 16.28 (2.80) | 0.876 | | 17.18 (2.99) | 14.53 (2.52) | **0.001** | ↗ |
| [ATP]$_{post}$ (mM) | 7.27 (1.25)* | 7.84 (0.82)* | 0.206 | | 7.77 (0.60) | 7.94 (0.62) | 0.312 | |
| [ADP]$_{post}$ (µM) | 54.85 (9.10) | 64.86 (28.22) | 0.282 | | 47.86 (8.79) | 55.87 (24.09) | 0.115 | |
| [H$_2$PO$_4^-$]$_{post}$ (mM) | 5.29 (1.09) | 5.31 (1.13) | 0.970 | | 5.51 (0.94) | 4.68 (0.92) | **0.002** | ↗ |
| pH$_{post}$ | 7.06 (0.04) | 7.07 (0.04)* | 0.395 | | 7.07 (0.03) | 7.07 (0.03) | 0.892 | |
| pH min | 6.93 (0.11)* | 6.88 (0.16) | 0.717 | | 6.97 (0.05) | 6.93 (0.12) | 0.190 | |
| pH change | 0.04 (0.03) | 0.02 (0.03)* | 0.351 | | 0.02 (0.02) | 0.02 (0.02) | 0.868 | |
| PCr depletion (%) | 39.51 (9.69) | 42.27 (14.84) | 0.629 | | 32.49 (7.45) | 37.41 (12.54) | 0.086 | |
| PCr change (mM) | 13.86 (3.34) | 15.87 (6.81) | 0.407 | | 13.26 (3.03) | 13.64 (5.22) | 0.744 | |
| $Vi_{PCr}$ (mM/s) | 0.31 (0.11)* | 0.50 (0.11) | **<0.001** | ↘ | 0.34 (0.10) | 0.42 (0.12) | **0.009** | ↘ |
| $V_{max}$ (mM/s) | 0.46 (0.13)* | 0.71 (0.12) | **<0.001** | ↘ | 0.52 (0.13) | 0.62 (0.16) | **0.009** | ↘ |
| QC outcomes | | | | | | | | |
| R2 $\tau_{PCr}Ex$ | 0.93 (0.03) | 0.93 (0.06)* | 0.442 | | 0.93 (0.03) | 0.93 (0.05) | 0.730 | |
| R2 $\tau_{Pi}Ex$ | 0.89 (0.08)* | 0.92 (0.06) | 0.545 | | 0.91 (0.05) | 0.91 (0.06) | 0.790 | |

| Recovery | COVID19 | | | | MS | | | |
|---|---|---|---|---|---|---|---|---|
| | Patients (N=15) | Controls (N=9) | P-values** | Variation | Patients (N=30) | Controls (N=32) | P-values** | Variation |
| $\tau_{PCr}Rec$ (s) | 43.33 (11.22) | 32.45 (10.66) | **0.029** | ↗ | 40.73 (9.29) | 33.11 (8.24) | **0.001** | ↗ |
| $\tau_{Pi}Rec$ (s) | 34.39 (7.74) | 27.83 (5.71)* | **0.025** | ↗ | 36.08 (10.49) | 28.55 (7.09) | **0.001** | ↗ |
| [PCr]$_{rec}$ (mM) | 37.78 (3.11) | 37.13 (2.76) | 0.608 | | 44.00 (5.76) | 36.58 (4.36) | **<0.001** | ↗ |
| [Pi]$_{rec}$ (mM) | 3.98 (0.87) | 3.25 (1.06) | 0.080 | | 4.01 (1.07) | 3.23 (0.84) | **0.002** | ↗ |
| [ATP]$_{rec}$ (mM) | 8.20 (0.12) | 8.20 (0.16) | 0.976 | | 8.22 (0.13) | 8.21 (0.14) | 0.789 | |
| [ADP]$_{rec}$ (µM) | 9.76 (0.86) | 9.44 (0.78) | 0.379 | | 9.58 (0.54) | 9.57 (1.07) | 0.946 | |
| [H$_2$PO$_4^-$]$_{rec}$ (mM) | 1.33 (0.33) | 1.12 (0.36) | 0.149 | | 1.36 (0.36) | 1.10 (0.28) | **0.003** | ↗ |
| pH$_{rec}$ | 7.05 (0.03) | 7.03 (0.03) | 0.120 | | 7.04 (0.02) | 7.04 (0.03) | 0.667 | |
| PCr repletion (%) | 48.57 (15.20)* | 43.51 (13.22) | 0.519 | | 36.64 (7.94) | 38.82 (11.09) | 0.381 | |
| PCr change (mM) | 18.30 (5.62) | 16.38 (5.71) | 0.429 | | 16.19 (4.63) | 14.25 (4.55) | 0.101 | |
| QC outcomes | | | | | | | | |
| R2 $\tau_{PCr}Rec$ | 0.93 (0.03) | 0.93 (0.08)* | 0.519 | | 0.93 (0.05) | 0.94 (0.05) | 0.691 | |
| R2 $\tau_{Pi}Rec$ | 0.88 (0.07) | 0.91 (0.05) | 0.269 | | 0.91 (0.06) | 0.91 (0.06) | 0.873 | |
| CV of $\tau_{PCr}Rec$ (%) | 5.37 (3.47) | 3.91 (1.97) | 0.263 | | 4.10 (2.18) | 5.00 (3.58) | 0.240 | |
| CV of $\tau_{Pi}Rec$ (%) | 5.97 (5.45)* | 6.33 (3.28) | 0.263 | | 4.03 (3.10) | 6.08 (2.67) | **0.007** | ↘ |

Table 3: COVID19, multiple sclerosis (MS) cohorts, and matched summary results with the Quality Control Score (QCS) and fixed T1 values. Results are descriptive statistics : mean (SD) . * The Shapiro-Wilk test has a value of p<0.05, the comparison between the populations is therefore made by a Mann-Whitney test with a statistical significance set at p<0.05. **T-test with Statistical significance was set to p<0.05. The variation indicates the direction of trends for the patient group compared to the control group. *** Fixed T1 values used as a correction factor, as proposed by the expert consensus[4].